# A Comparative Analysis of XML Documents, XML Enabled Databases and Native XML Databases

Amir Mohammad Saba[1], Elham Shahab[1], Hadi Abdolrahimpour[2], Mahsa Hakimi[3], Akbar Moazzam[4]
Department of Computer Engineering, Azad Islamic University, Yazd, Iran[1]
{amsaba, ma.shahab}@iauyazd.ac.ir
Department of Bio Mechanic, Azad Islamic University, Yazd, Iran[2]
Department of Pharmacy, Shahid Beheshti University, Tehran, Iran[3]
Department of Computer Science, Isfahan University, Isfahan, Iran[4]

*Abstract*— With the increasing popularity of XML data and a great need for a database management system able to store, retrieve and manipulate XML-based data in an efficient manner, database research communities and software industries have tried to respond to this requirement. XML-enabled database and native XML database are two approaches that have been proposed to address this challenge. These two approaches are a legacy database systems which are extended to store, retrieve and manipulate XML-based data. The major objective of this paper is to explore and compare between the two approaches and reach to some criteria to have a suitable guideline to select the best approach in each circumstance. In general, native XML database systems have more ability in comparison with XML-enabled database system for managing XML-based data.

## I. INTRODUCTION

Nowadays, with remarkable expansion of the information on the internet, one of the basic challenge is that how we can exchange information between different types of applications and e-business independently. eXtensible Markup Language(XML) standard promises to be the standard way for data exchanging in e-business and applications.

XML which is known as a mark up language and textual file format propose a way for a description of a document content with non-predefined tags. It utilizes tags called elements to describe the data enclosed within elements. With the ability to create custom tags to describe data, XML has quickly spread and played an important role as one of a major means for data exchange [3]. In fact, Heterogeneity of data records, extensibility by allowing different data types in single documents beside the flexibility in size are number of advantages of using XML-based documents in comparison with other approaches that make it completely different from other ones.

In recent years, with the increase usage of XML structure and growing size of XML documents collections it is necessary to adopt efficient techniques to store and retrieve data from XML-based documents management systems. Figure 1 follows the evaluation of traditional data management systems to about current systems. Storing XML-based documents has presented a challenge for commercial Database Management Systems (DBMS) because they were not able to naturally preserve the conceptual design of XML data models and required data manipulation and transformation before storing XML structures[2].

Object Oriented databases that could preserve XML data structures as objects could not compete with speed and reliability with their DBMS commercial competitors [3]. However, as the importance of and demand for storing XML data increased and different techniques for data extraction and inference from XML–based through applying summarization [13] and knowledge extraction are growing [7,12], two new approaches which are XML Enabled Database and Native XML Database have emerged to address those [2].

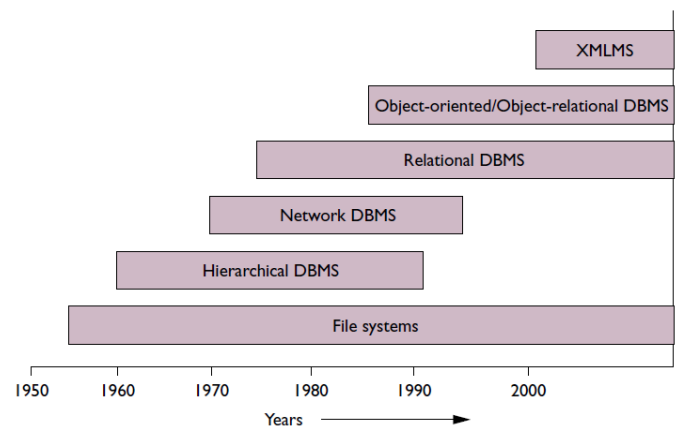

Figure 1: Evaluation of traditional data management systems [3].

Using XML-enabled or native XML database in many cases is based on specific application and the format of XML documents. Before explanation of each approach we have a quick glance to some terminology in XML documents.

## II. XML DOCUMENTS

Data-centric and documents-centric are two types of XML documents[1]. In data-centric documents, data are well-structured, appear in a regular order and relatively follow fixed XML schema. These documents are used for communicating date between companies or applications. For examples following XML document is data-centric. It is clearly well-structured and contains no mixed content. Other examples like sales orders, scientific data and stock quotes are considered as data-centric documents.

<Memo>
<Conference date="05/10/2017" time="08:00AM">IEEE Conference</ Conference>
< Purpose>Attending the Conference</ Purpose >
<Location>Conference Hall </Location>
</Memo>

On the other hand, documents-centric XML documents are those ones that are loosely structured, do not follow fixed XML schema and usually are designed for human consumption. For example the following memo document is document-centric.

<Memo>
Please Computer Science members come to
<Location> Conference Hall </Location>on
<ConferenceDate>05/10/2017</ConferenceDate > at
<ConferenceTime>08:00AM</ConferenceTime > to
<Purpose>Attending the Conference</Purpose>
</Memo>

Querying data in XML documents is done through XQuery which is a declarative, typed, functional language design for querying data stored in XML format. XPath is a specialized expression language used to parse through XML hierarchies and retrieve chunk of data from XML documents [3].

## III. XML AND DATABASE

Software and data management professionals are almost always encountered with some challenges in regards of managing data between XML and relational data structures [4]. Although, technological advances have made some basic approaches to address these data managements challenges but software developers or data managers should consider their particular circumstance and weigh them against the proposed options and approaches. In this way, there are two basic strategies to manage data delivered in XML documents. The first one which is known as XML enabled database is implemented through mapping the document's data model to database model and converting XML data into database based on that mapping[3]. The second method which is called native XML database is mapping XML model into a fixed set database structure[3].

## IV. XML ENABLED DATABASE

XML enabled database is one of the latest development to address the challenges associated with XML data and relational databases. The unit of storage in relational database systems is table row which is known as record and the best data to be stored in that database system is well-structured XML documents. Figure 2 shows an overview of how XML enabled database works during processing the XML document.

At the first step, at the XQuery interface, an XML Query is received from Entry point and then validate for correctness [6]. After the first step, XML Query comes for XML schema validation based on the method of storage and then parses into various parts of hierarchical structure by Document Object Model (DOM) or Simple API for XML (SAX) for parsing an XML document. In the next step, XQuery/SQL interface plays an important role in XML document processing to convert the output XQuery from last step and parse it into set of SQL statement that can be applied to the database. In this step two scenario can be used to store XML data in a relational database system. In this context, the entire XML document is fed into database as an entry into a column a row whose data type is specified as Large Object (LOB).

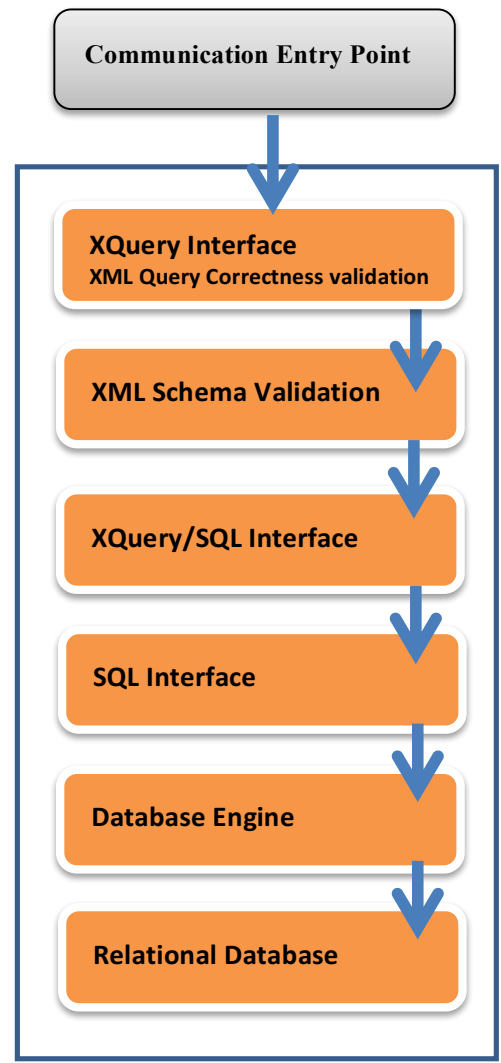

Figure 2: XML document process in a typical XML Enabled Database (derived from [3]).

The second story in regards to storing data in an XML database lies in mapping the content of the document to a relational database structure. This process of conversation is called shredding. In this step the XML document must be validate against XML schema that it can be generated manually or automatically through adaptive XML shredding.

In general, the entire processing of XML document is done through a layer that is invisible from the relational database system.

On the other side, similar processing is required in reverse side when there is a requested data from the database which is known as XML publishing.

## V. NATIVE XML DATABASE

Another approach to basic challenges between XML data and relational database is native XML database. Native XML database is the database that stores XML documents directly. The main difference between this approach and relational database system is the fundamental unite of storage which is XML documents in Native XML database and row or record in relational database systems. Collection is set of related XML documents in native XML database that plays a similar role like a table in relational database system.

In a native XML database, XML is visible inside the database and there is a unique database for all XML schemas and documents. Native XML databases are especially suitable for storing irregular, deeply hierarchical and recursive data[4]. Common native XML database applications include document management, support for semi-structured data, support for sparse data, catalog data, manufacturing parts databases, medical information storage and etc[6].

A typical process for native XML database is shown in figure 3. At the first XML document pass the Communication Entry Point and in XQuery interface is passed using SAX or DOM and then in the next step in the XML Schema Validation is validated against XML schema. As it is depicted in figure 3 the validation stage is an optional stage but it is a good practice to ensure that multiple documents conform to one another[3]. The schema can be used as a documentation of what element exists in database so each XML document can be stored with its own schema. This facility in native XML database approach deliver a good flexibility in data storage when new elements needs to be stored in the database. This is basically supported by the fact that XML documents naturally support document, element order, and sibling order[3].

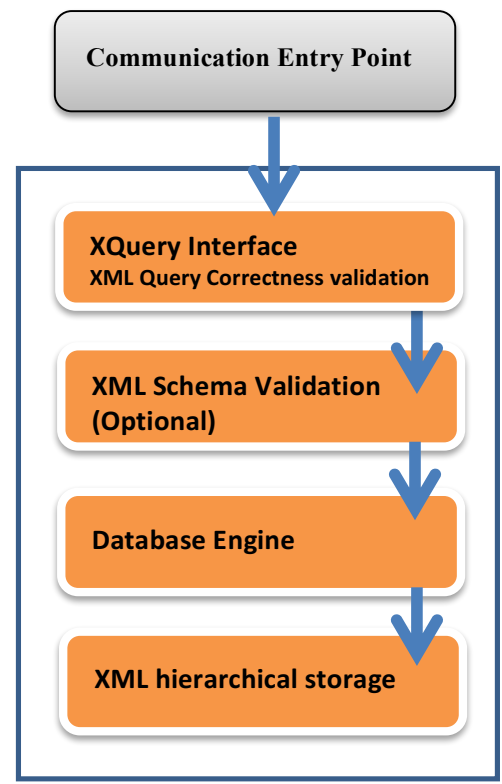

Figure 3: XML document process for a typical Native XML Database (derived from [3]).

The data which are retrieved from an XML document are stored directly into hierarchical tree structure which is completely the reflection of the default nature of XML documents.

## VI. DISCUSSING

Based on figure 2 and figure 3, it is clear that there are much more processing steps in XML enabled databases in comparison with native XML databases. This difference can address the performance level of each method. In fact, the most valuable processing time has to be spent on :
- Mapping XML schemas to database structure.
- Translating XQuery into SQL.
- Parsing and reparsing XML documents stored in LOB.

Which are completely accomplished in XML enabled databases. Based on [3], the main reason in regard of processor computing is parsing XML document so it is recommended that XML documents must be heavily indexed to increase the performance before parsing [3]. In XML enabled databases, adaptive shredding adds more processing overhead[2], so basically using manually XML schema can decrease some overheads in this approach. In general, relational databases are unable to perform full text search without loading the entire document into memory[4] but XML aware system makes this facilities for native XML databases to search XML documents while they have been partially loading into memory. Table 1 shows the basic criteria in regards of two approaches.

Table 1: comparison native XML and XML enabled approaches (derived from [1,4,5,9,10,11]).

| Criteria | Native XML | XML Enabled |
|---|---|---|
| **Access to specific data element** | Weak | Very good |
| **Maintaining the hierarchical XML data** | Good | Very good |
| **Storing complete XML document** | Good | Weak |
| **Lost of document identity** | Good | Weak |
| **Irregular XML data structure** | Very good | Good |
| **Managing collection of semi-structure XML** | Very good | Good |
| **XML with no schema** | Very good | Weak |
| **XML document centric date** | Very good | Weak |

## VII. CONCLUSION

Although, relational database systems are very popular to use but the processing of XML documents and in particular semi-structured XML documents which account for the majority, is the area that we have to take two main approaches into our consideration [3]. XML enabled databases are a good solution when it comes to XML data exchange between applications and other databases[1]. An important consequence of using XML as a data exchange format is that an XML enabled database will only retain information captured by the underlying data model. Moreover, this model is not the most appropriate for fully and efficiently managing large XML documents, as shredding/publishing XML data to/from relational tables requires a large number of join operations. Native XML databases are focused on applications that need to store whole XML documents, in contrast with applications that simply store data that may or may not have been in an XML format[1]. The two fundamental reasons to choose native XML database is performance and space. Depending on the storing strategy adopted, a native XML database may have much better performance in retrieving and/or recomposing documents, in structural queries and full-text searches[3]. If performance is not an important issue, the second issue to consider is wasted space: the irregular structure of document-centric XML documents will unavoidably lead to a large number of null values if we use XML enabled databases. Studies have also shown that in reality native XML databases have better performance in comparison with XML enabled databases approaches. We think with enough research and development of native XML databases, they have enough potential to become the dominant database systems to process XML-based documents.